\documentclass[superscriptaddress,showpacs,amssymb,aps,amsmath,twocolumn]{revtex4}
\usepackage{epsfig}

\newcommand{\upd}{{\rm d}}
\newcommand{\be}{\begin{equation}}
\newcommand{\ee}{\end{equation}}

\begin{document}

\title{Real space origin of temperature crossovers in supercooled
liquids}

\author{Ludovic Berthier}

\affiliation{Theoretical Physics, University of Oxford, 1 Keble Road,
Oxford, OX1 3NP, UK}

\affiliation{Laboratoire des Verres, Universit\'e Montpellier II,
34095 Montpellier, France}

\author{Juan P. Garrahan}

\affiliation{Theoretical Physics, University of Oxford, 1 Keble Road,
Oxford, OX1 3NP, UK}

\date{\today}

\begin{abstract}
We show that the various crossovers between dynamical regimes observed
in experiments and simulations of supercooled liquids can be explained
in simple terms from the existence and statistical properties of
dynamical heterogeneities. 
We confirm that dynamic heterogeneity is
responsible for the slowing down of glass formers at temperatures well
above the dynamic singularity $T_c$ predicted by mode coupling theory.
Our results imply that activated processes govern
the long-time dynamics even in the temperature regime 
where they are neglected by mode-coupling theory.
We show that alternative interpretations
based on topographic properties of the potential energy landscape
are complicated and inefficient ways of describing simple
physical features which are naturally accounted for within our approach.  
We show in particular that the reported links
between mode coupling and landscape singularities do not exist.
\end{abstract}



\maketitle

\section{Introduction}
\label{introduction}

The aim of this paper is to critically reconsider the physical origin
of the onset of dynamical arrest and the associated crossovers between
distinct dynamical regimes displayed by liquids supercooled through
their melting temperature towards the glass 
transition~\cite{Ediger-et-al,Angell,Debenedetti-Stillinger,Debenedetti}.
We do this by extending the real space theoretical framework based on
dynamic facilitation of Refs.\
\cite{Garrahan-Chandler,Berthier-Garrahan,Garrahan-Chandler2,Berthier} to the
moderately supercooled regime corresponding to the region
where mode coupling theory (MCT)~\cite{MCT} supposedly applies, as
reviewed in \cite{MCT-experiments,MCT-simulations}.
This novel approach takes directly into
account the spatial aspects of the dynamics, in particular those
related to dynamic heterogeneity \cite{DHReviews}, in contrast with many
other theories \cite{Debenedetti-Stillinger,Debenedetti,MCT}.
Our analysis
shows that the onset of slowing down can be understood in a simple
physical way in terms of the dynamical properties of effective
excitations, or defects, as a progressive crossover from a regime of
fast dynamics dense in defect clusters, to one of slow heterogeneous
dynamics dominated by isolated localized defects. 
We demonstrate
that this real space picture explains the
observed crossover temperatures, challenges the idea that these
crossovers are related to changes in the topography of the energy
surface or to MCT singularities, and is able to account for the
apparent correlations observed between ``landscape'' and dynamical
properties.

The paper is organized as follows. In the rest of the Introduction we
review the MCT and energy landscape points of view, discuss their
problems and limitations, and describe the alternative real space
perspective we will pursue.  In Section \ref{physics} we
develop the physical picture of the onset of slowing down and
dynamical crossovers which emerges from our theoretical approach.  In
Section \ref{consequence} we discuss its quantitative consequences and
compare them to published numerical results.  In Section
\ref{landscape} we show how our approach also enables to derive 
the observed properties of the potential energy landscape of
supercooled liquids.  Finally, in Section
\ref{conclusion} we discuss our results and state our conclusions.

\subsection{MCT/landscape scenario...} 

It is often assumed that the initial slowing down of the dynamics of
supercooled liquids can be rationalized by MCT
\cite{Ediger-et-al,Angell,Debenedetti-Stillinger,Debenedetti,MCT-experiments,MCT-simulations}.  
Numerical simulations are now able to
investigate the first five decades in time of this slowing down
\cite{MCT-simulations}, so this is also the regime which has been
studied in greatest microscopic detail. The degree of success of MCT
is still a matter of debate.  This is due to the fact that the central
MCT prediction, a complete dynamical arrest at a temperature $T_c$
where the alpha-relaxation time diverges as a power law,
$\tau_\alpha(T) \sim (T-T_c)^{-\gamma}$, is actually never observed,
but a power law fit to the data apparently works on a restricted time
window \cite{MCT-experiments,MCT-simulations}.  
The appearance of new mechanisms for
relaxation, often termed ``activated processes'', but seldom described
in any detail, is then invoked to explain the discrepancy between
observations and MCT predictions. In fact, activated processes are
actually quantitatively defined, within MCT,
by deviations between data and
predictions~\cite{MCT-simulations,Kob-Andersen,thomas}.
It is believed that activated processes become relevant close 
to the dynamical singularity $T_c$,
their main effect being to prevent the predicted transition.

From $T_c$ downwards it is assumed that the physics is dominated by
activated processes, which determine also the canonical features of glass
transition phenomena \cite{kivtar2}: non-exponential relaxation,
strong and fragile liquid behaviours, decoupling between transport
coefficients, etc.  It is sometimes said that the relevant physics for
the glass transition sets in at $T_c$, and is therefore out of reach
of numerical simulations \cite{Donth}.  Crossovers into the activated
dynamics regime are also reported to occur at temperatures $T_x$
\cite{Rossler-Sokolov} or $T_B$ \cite{Stickel-et-al}, depending on
which aspect of the physics is considered.  It is believed that all
these temperatures are close enough to be taken as equivalent, $T_c
\approx T_x \approx T_B$ \cite{Angell-10Q}.

\begin{figure}[ht]
\begin{center}
\epsfig{file=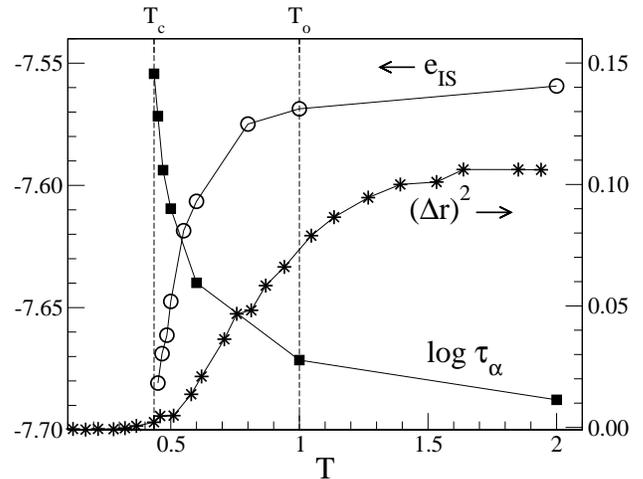, width=8.25cm}
\caption{\label{crossoverLJ} Onset of slowing
down in the binary Lennard-Jones mixture of Ref.\
\cite{Kob-Andersen}. Three quantities are reported as a function of
temperature $T$.  (i) Logarithm of the relaxation time, $\log
\tau_\alpha$ (in arbitrary units extending over four decades in time);
a MCT power law divergence, with $T_c =0.435$, was used to fit this
data in Ref.\ \cite{Kob-Andersen}.  (ii) Energy of inherent
structures, $e_{IS}$, taken from Ref.\ \cite{kobsciortino}, which
decreases markedly when the temperature decreases below $T_o = 1.0$. 
(iii) Anharmonic part of Cartesian distance between
configurations and their corresponding IS, $(\Delta r)^2 \equiv N^{-1}
\sum_i ({\bf r}_i-{\bf r}_i^{\rm (IS)})^2 - a~T$, which displays a
qualitative change around $T_c$, taken from Ref.\
\cite{Sastry-et-al}.}
\end{center}
\end{figure}

The above scenario is apparently corroborated by the study of the
statistical properties of the potential energy landscape of model
liquids
\cite{Goldstein,Stillinger-Weber,Stillinger,Debenedetti-Stillinger}.
From the properties of the landscape two temperatures seem to emerge,
$T_o$ and $T_c$ \cite{Sastry-et-al}.  The onset of slowing down of the
dynamics takes place at $T_o$, and coincides with the temperature
below which the average energy of inherent structures (IS), i.e.,
local minima of the potential energy \cite{Stillinger-Weber},
$e_{IS}(T)$, starts to decrease markedly, 
see Fig.~\ref{crossoverLJ}. This has been interpreted
as the sign that the landscape starts to ``influence'' the dynamical
behaviour \cite{Sastry-et-al}.  At $T_c$, it is further argued, a second
change in the landscape properties takes place, which is indicated by several 
observations~\cite{Sastry-et-al,Buchner-Heuer1,Buchner-Heuer2,schroder,jund,keyes1,boson}.
For example, the mean-square displacement from an equilibrated
configuration to its corresponding inherent structure, $N^{-1} \sum_i
({\bf r}_i-{\bf r}_i^{\rm (IS)})^2$, is proportional to $T$ below
$T_c$, as expected from pure vibrations in quadratic wells, but the
temperature dependence changes above $T_c$, revealing
``anharmonicities'' in the landscape
\cite{Sastry-et-al,Buchner-Heuer1}, see Fig.~\ref{crossoverLJ}.

Another indication of a topological change in the energy landscape was
discussed in Ref.\ \cite{Cavagna}, in analogy with what happens in
mean field models \cite{Kurchan-Laloux,Cavagna-et-al}: the vanishing
as $T$ approaches $T_c$ of the mean intensive number of negative
directions (intensive index) of stationary points of the potential
energy, $n_s(T)$.  Numerical simulations
\cite{Broderix-et-al,Angelani-et-al,saddle2,doye,stariolo,dh2} found that
$n_s(T)$ decreases with decreasing $T$, and fits were performed to
show that $n_s(T_c)=0$ \cite{Broderix-et-al,Angelani-et-al,saddle2}.
The physical interpretation of this result is the 
apparent existence at $T_c$
of a ``geometric transition'' between a ``saddle dominated regime''
above $T_c$ and a ``minima dominated regime'' below $T_c$
\cite{Cavagna2}.  $T_c$ would then really coincide with the appearance of
activated processes, described in a topographic language as
``hopping'' between minima of the landscape.  Analogous findings had
been previously reported from instantaneous normal mode analysis of
equilibrium configurations \cite{laird}, also supporting a qualitative
change in the landscape topology close to $T_c$ \cite{INM,keyes2}.

Figure \ref{crossoverLJ} summarizes this MCT/landscape scenario
with published numerical data~\cite{Kob-Andersen,kobsciortino,Sastry-et-al} 
obtained for the standard
supercooled liquid model of Ref.~\cite{Kob-Andersen}.

\subsection{... and its problems and contradictions}

At first sight Fig.\ \ref{crossoverLJ} appears as convincing evidence
in favour of the MCT/landscape interpretation of the dynamics.
On closer inspection, however, the above scenario is less robust.
Several qualitative and quantitative observations do not fit into the
picture presented above.

\textit{(i) Activated dynamics above $T_c$.}  The main idea behind the
landscape approach \cite{Goldstein} is that vibrations and structural
relaxations take place on very different time scales, so that the
system is ``trapped'' and vibrates in one minimum before ``hopping''
to another minimum.  This is indeed observed using the mapping from
trajectories to IS \cite{Stillinger-Weber} in simulations of
sufficiently small systems \cite{Doliwa-Heuer,Denny-et-al}.  We have
also discussed theoretically this issue in a recent work
\cite{Berthier-Garrahan}.  Given that numerical studies were performed
much above $T_c$, a crucial conclusion of Refs.\
\cite{Doliwa-Heuer,Denny-et-al,Berthier-Garrahan} is that ``activated
dynamics'' is indeed present in this temperature regime.

\textit{(ii) Heterogeneous dynamics above $T_c$.}  It is now
well-documented that the dynamics of supercooled liquids, even above
the mode coupling temperature $T_c$, is heterogeneous in the sense
that the local relaxation time has non-trivial spatial correlations
\cite{Glotzer}.  This phenomenon is not very different from what
happens experimentally close to the glass transition $T_g$
\cite{DHReviews}.  Besides, the decoupling between transport
coefficients, which is also interpreted in terms of dynamical
heterogeneity \cite{DHReviews,kivtar3}, is observed in numerical simulations
above $T_c$ \cite{Kob-Andersen}, although the effect is 
quantitatively less pronounced than in experiments near 
$T_g$~\cite{gilles1}. 

\textit{(iii) Presence of saddles below $T_c$.}  Despite claims based
on numerical results that $T_c$ marks a real change in the topology of
the landscape \cite{Angelani-et-al,Broderix-et-al,saddle2}, there are strong
indications that this is at best only a crossover \cite{doye,stariolo}, and at
worst a biased interpretation of numerical data~\cite{dh2}.  For
instance, careful numerical studies have shown that the saddle index
$n_s(T)$ remains positive even below $T_c$ \cite{doye}.  Moreover,
Ref.\ \cite{dh2} argues convincingly that the data for $n_s(T)$ can be
described by an Arrhenius law, $n_s(T) \sim \exp (-E/T)$, with $E$ an
energy scale, which means that 
$T_c$ does not mark any particular change
in the saddle index. 

\subsection{Alternative: real space physics and coarse-grained 
models}

The problems
described above can be overcome through
an alternative perspective on glass
transition phenomena which puts the real space aspects of
the dynamics at its core.  This is the approach developed in Refs.\
\cite{Garrahan-Chandler,Berthier-Garrahan,Garrahan-Chandler2,Berthier}.
Interestingly, several of its central concepts, like the relevance to
the dynamics of localized excitations \cite{Glarum,Anderson} and the
importance of effective kinetic constraints
\cite{Fredrickson-Andersen,Palmer-et-al}, have been present in
the literature for many years.  Moreover, 
one of the original key observations
of dynamic heterogeneity in glass formers was made by Harrowell and
coworkers \cite{Harrowell} in the models of Ref.\
\cite{Fredrickson-Andersen}.  See Ref.\ \cite{Ritort-Sollich} for an
exhaustive review.

Our approach relies on only two basic observations.
(i) At low temperature
mobility within a supercooled liquid is sparse and very few particles
are mobile.  This is somewhat equivalent to the statement that
particles are ``caged'' for long period of times, as reflected by a
plateau in the mean-square displacement of individual particles.
(ii) When a microscopic region of space is mobile
it influences the dynamics of neighbouring regions, enabling them to
become mobile, and thus allowing mobility to propagate in the system.
This is the concept of dynamic facilitation
\cite{Anderson,Fredrickson-Andersen}.  The observation that very
mobile particles in a supercooled liquid move along correlated
``strings'' \cite{Glotzer} is a confirmation of this fundamental idea.

From these two concepts it is possible to build effective microscopic
models for glass formers by means of a coarse-graining procedure.
This procedure can be schematically described as follows
\cite{Garrahan-Chandler2}.  Spatially, the particles are
coarse-grained over a length scale $\delta x$ of the order of the
static correlation length given by the pair correlation function.
This removes any static correlations between coarse grained regions of
linear size $\delta x$.  Cells are then identified according to their
mobility by performing a coarse-graining on a microscopic time scale
$\delta t$.  In its simplest version cells are identified by a scalar
``mobility field'', $n({\bf r},t)=0,1$, the values $0/1$ corresponding
to an immobile/mobile cell at position ${\bf r}$ and time $t$.  The
next step is to replace continuous space by a lattice, $n({\bf r},t)
\to n_i(t)$.  Mobile or excited cells carry a free energy cost, so
when mobility is low it is reasonable to describe their static
properties with a non-interacting Hamiltonian~\cite{Fredrickson-Andersen},
\begin{equation}
H = \sum_{i=1}^N n_i ,
\label{ham}
\end{equation} 
for a lattice of $N$ sites.  The link between mobility and potential
energy \cite{Garrahan-Chandler} has also been observed in numerical
simulations~\cite{Glotzer}.

The coarse-graining procedure described above will generate local
dynamical rules for the mobility field.  The prominent feature of this
dynamics will be dynamic facilitation, which in its simplest version
states that a cell at site $i$ is allowed to move only if it has an
excited nearest neighbour \cite{Fredrickson-Andersen},
\begin{equation}
n_i=0 
\begin{array}{c}
\xrightarrow{~ ~ ~ {\cal C}_i ~ c ~ ~ } \\ \xleftarrow[{\cal C}_i ~
(1-c)]{} \\
\end{array}
n_i=1 ,
\label{rates}
\end{equation}
where ${\cal C}_i = 1 - \prod_{\langle j,i \rangle} (1-n_j)$, and
$\langle j,i \rangle$ indicates nearest neighbour, and $c$ represents
the average concentration of excited cells easily deduced 
from (\ref{ham}):
\begin{equation}
c(T) \equiv \langle n_i \rangle = (1+e^{1/T})^{-1}.
\label{ct}
\end{equation}
Explicit examples where dynamic facilitation is generated under
coarse-graining can be found in Ref.~\cite{Garrahan-Newman}. 
Clearly, different models are defined simply
by changing the kinetic rules, e.g., the number or directionality of
mobile neighbours required to move \cite{Ritort-Sollich}.  Also, a
more complex mobility field may be required to account quantitatively
for all glass transition features \cite{Garrahan-Chandler2}.

Crucially, we will show that the physical mechanisms which explain the
onset of slowing down and crossovers between different dynamical
regimes in supercooled liquids are generic to this class of models.
This means that we can use the simplest of them, the Fredrickson-Andersen
(FA) model defined by Eqs.\ (\ref{ham},\ref{rates}) in one spatial
dimension (hereafter 1D FA model) 
to make detailed predictions and calculations.

\section{Physical picture of dynamic crossovers}
\label{physics}

\begin{figure}
\epsfig{file=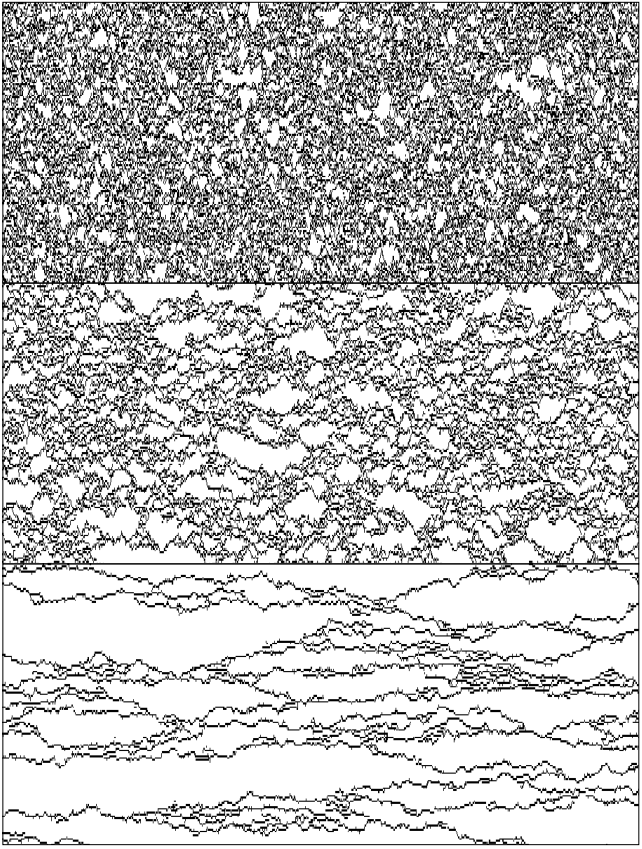,width=8.5cm}
\caption{\label{traj} Representative trajectories in the 1D FA model.
The vertical axis is space, the horizontal one time.  The three
trajectories are for $L=150$ and $t=2000$.  Excited cells (or defects)
are black, unexcited ones white.  The top frame is for $T=2.5$, in the
high temperature regime where almost no isolated defects are present.
The middle frame is for $T=1.0$, the temperature regime where slow
bubbles start to appear, seen here as large white domains.  The bottom
frame is for $T=0.5$, where almost all defects are isolated.  For
$T=0.5$, the mean relaxation time is $\sim 120$, the mean dynamic
correlation length $\sim 9$, but it is clear that times and lengths
are broadly distributed.}
\end{figure}

In order to understand the physics captured by the coarse-grained
facilitated models defined above, it is useful to look at
trajectories, that is, space-time representations of the dynamics
\cite{Garrahan-Chandler}.  We show in Fig.\ \ref{traj} three
representative trajectories for the 1D FA model, where mobile cells
(defects) are black, and immobile ones are white.  From the
trajectories, the principal observation is the appearance at low
temperatures of non-trivial spatio-temporal correlations, seen as
spatially and temporally extended domains of immobile cells delimited
by isolated defects \cite{Garrahan-Chandler}.  In 1D they look like
``bubbles''~\cite{Berthier-Garrahan}, 
and trajectories are dense assemblies of these slow
bubbles.  This nanoscopic ordering in trajectory space is the cause of
the phenomenon of dynamic heterogeneity
observed experimentally and in simulations~\cite{Garrahan-Chandler}.  
Dynamic heterogeneity is the central aspect of
the physics of supercooled liquids: it is
naturally captured by our approach.

The statistical mechanics of trajectories, rather than configurations,
determines the dynamical behaviour.  For example, due to the
non-interacting Hamiltonian (\ref{ham}), static correlations are
trivial.  However, when trajectories are considered, it is clear that
cells become dynamically correlated.  In other words, these models
naturally predict the existence of a dynamical correlation $\ell(T)$
which grows when the dynamics slows down.  This statement can be
quantified \cite{Garrahan-Chandler,Berthier,silvio,doliwa} by studying
multi-point functions, for example, $C(|i-j|,t)= \langle P_i(t) P_j(t)
\rangle - \langle P_i(t) \rangle \langle P_j(t) \rangle$, where
$P_i(t)$ is a dynamical correlator at site $i$ (below we will consider
the persistence of site $i$).  The spatial decay of a function like
$C(|i-j|,t)$ defines unambiguously the dynamical correlation length
$\ell(T)$, as already discussed theoretically
\cite{Garrahan-Chandler,Berthier,silvio} and measured in numerical
simulations \cite{silvio,Glotzer,doliwa,harro2,length}. Furthermore, the joint
distributions of time and length scales give rise to the canonical
features of glass formers, like stretched relaxation, decoupling
between transport coefficients, and (kinetic and thermodynamic) strong
and fragile 
behaviours~\cite{Garrahan-Chandler,Berthier-Garrahan,Garrahan-Chandler2}.

Let us take a closer look at the temperature evolution of the
trajectories in Fig.\ \ref{traj}. Starting from the very low
temperatures where trajectories consist of a mixture of slow bubbles,
the dynamics can be understood in terms of the opening and closing of
bubbles, that is, the branching of an excitation line 
or the coalescence of two.
As shown in \cite{Berthier-Garrahan},
these events are the ``hopping between
minima'' described in Ref.\ \cite{Goldstein}. 
Therefore, we have a clear understanding of
``activated processes'' and of their
statistical properties~\cite{Berthier-Garrahan}.

As temperature is increased more and more defects are present.  This
has several consequences.  First, the typical spatial and temporal
extension of bubbles reduces, that is, the system becomes faster and
less heterogeneous.  Second, clusters of defects become more common.
These objects are important because their dynamics is completely
different from that of isolated defects.  In a cluster, defects do not
have to diffuse and create or annihilate other defects but can
instantaneously relax in a much faster process.  At high temperature, the
dynamics is fast because almost no bubbles are present, and the
dynamics is governed by clusters of defects.  Interestingly, at some
intermediate temperature (middle frame in Fig.\ \ref{traj}) a
coexistence between clusters and isolated defects is observed, so that
the dynamics has a ``mixed'' character.

Clusters of defects disappear much faster with
decreasing temperature than the overall concentration of defects.  The
probability to have a cluster of $k$ defects is 
indeed $p(k) \propto c^k$, so
that at low $T$ we have $p(1) \gg p(2) \gg \cdots$.

\begin{figure}
\epsfig{file=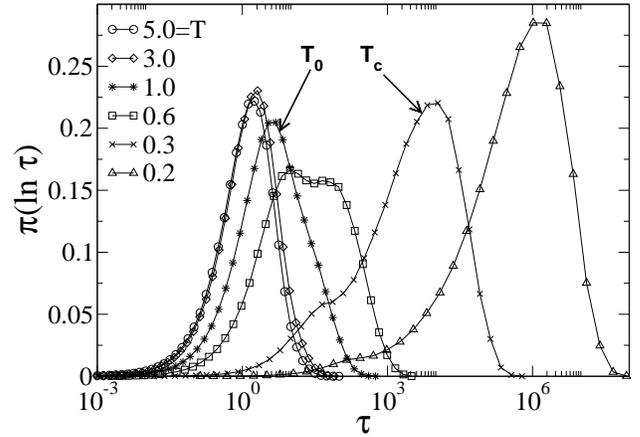,width=8.25cm}
\caption{\label{distrib} Distribution of the logarithm of the
persistence time of individual cells, $\pi(\ln \tau)$ for various
temperatures.  $T_o=1.0$ marks the appearance of a shoulder in the
high temperature distribution. For $T_c < T=0.6 < T_o$, two
``processes'' coexist. Fast processes disappear close to $T_c=0.3$.}
\end{figure}

The coexistence of fast and slow processes with different temperature
behaviour has a direct influence on the distribution of local
relaxation times, which we present in Fig.\ \ref{distrib} for various
temperatures.  At high temperature, $T \gg 1.0$, where fast processes
are dominant, the distribution is exponential, with a mean which
depends weakly on temperature (below we discuss in detail its
temperature dependence).  Around $T=1.0$, a shoulder develops in the
large time tail of the distribution, corresponding to the appearance
of the bubbles in the trajectories of Fig.\ \ref{traj}.  This marks
the increasing relevance of slow processes and the growth of the
dynamic correlation length beyond the microscopic high temperature
value.  This temperature corresponds 
therefore to the onset temperature, $T_o =
1.0$.

Decreasing further the temperature, we clearly see a regime of mixed
dynamics.  At $T=0.6$, for instance, there are two peaks in the
distribution, reflecting the coexistence of clusters and isolated
defects.  At this temperature, the time scale has already increased by
several orders of magnitude, and the dynamic correlation length is
about $\ell(T=0.6) \approx c^{-1}(T=0.6) \approx 6$.

Finally, further decrease in temperature makes clusters of defects
very rare and we are left only with the contribution of isolated
defects.  This low temperature distribution is the one discussed in
Refs.\ \cite{Garrahan-Chandler,Berthier-Garrahan}, which in turn
implies the stretched exponential decay of dynamical correlators.  The
contribution of clusters becomes negligible beyond a second
crossover temperature, here $T_c=0.3$.  While this crossover
temperature is not linked in any way to the mode coupling singularity
$T_c$, this choice of notation will become clear shortly.

\begin{figure}
\epsfig{file=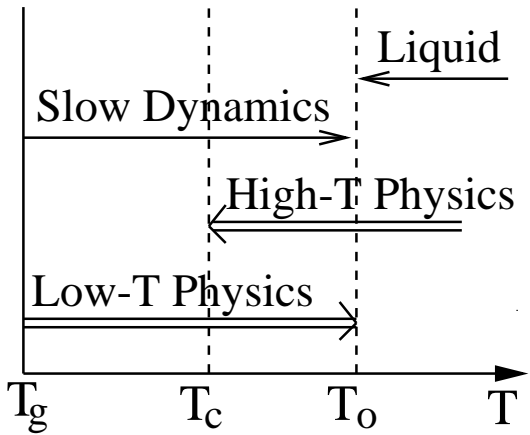,width=7.cm}
\caption{\label{regime} Temperature regimes emerging from the
discussion of section~\ref{physics}.  $T_o$ marks the onset of slow
dynamics, the appearance of the isolated defects (bubbles, activated
processes), and the growth of a dynamic correlation length.  At $T_c$,
traces of the high-$T$ physics (clusters) become negligible in the
distributions of relaxation times.  The crossover region $T_c < T <
T_o$ has therefore a mixed character.}
\end{figure}

From these distributions, it is possible to propose an empirical but
quantitative determination of $T_o$ and $T_c$. At each
temperature $T$, the distribution is composed of fast processes, $\tau
< \tau_\star(T)$, and slow processes, $\tau > \tau_\star(T)$, 
where $\tau^\star (T)$ can be defined, e.g., as in Ref.~\cite{heuer}.
Requiring that slow processes are a significant fraction (say 90\%) of
the distribution leads to the definition of $T_c$:
\begin{equation} 
\int_{\tau_\star(T_c)}^\infty \upd \tau' \pi(\tau') = 0.9,
\label{Tc}
\end{equation}
where in an abuse of notation we also call $\pi(\tau)$ the distribution of
persistence times.  Requiring that slow processes, while not dominant
in number, still contribute to a significant fraction of the mean
relaxation time (say again 90\%) leads to the definition of $T_o$:
\begin{equation} 
\int_{\tau_\star(T_o)}^\infty \upd \tau' \pi(\tau') \tau' = 0.9 ~
\langle \tau \rangle .
\label{To}
\end{equation}

In Fig.\ \ref{regime}, we summarize the physical picture that emerges
from the considerations of this section.  From the distributions of
Fig.\ \ref{distrib}, we recognize that the dynamics becomes slow when
the temperature is decreased below the onset temperature $T_o$.  This
distinguishes the trivial liquid and the slow glassy regimes.  From
the trajectories of Fig.\ \ref{traj}, we were also able to distinguish
between fast, non-activated processes (clusters), and slow, activated
processes (isolated defects, bubbles).  The former, typical of the
liquid high-$T$ physics, become negligible below $T_c$, while the
latter, typical of the low-$T$ physics, appear at the onset
temperature $T_o$.  As a consequence, the crossover region $T_c < T <
T_o$ contains traces of both high-$T$ and low-$T$ physics, as observed
in the time distributions of Fig.\ \ref{distrib}.

\section{Quantitative consequences}
\label{consequence}

The physical picture we have presented for the onset of slowing
down, based on the increasing relevance of a dynamically heterogeneous
evolution of the system, leads to quantitative predictions which are
in good agreement with previous numerical and experimental studies,
as we discuss in this section.

\subsection{Dynamical correlators}

The basic dynamical quantities recorded in experiments and simulations
of supercooled liquids are spatially averaged two-time 
functions.  Simulations usually focus on the time domain and typically
consider density-density correlation functions, while experimental
results are often expressed in the frequency domain, measuring for
instance dielectric susceptibilities.  We will only consider systems
in equilibrium so that the information content of both kinds of
measurements is equivalent.

\begin{figure}
\epsfig{file=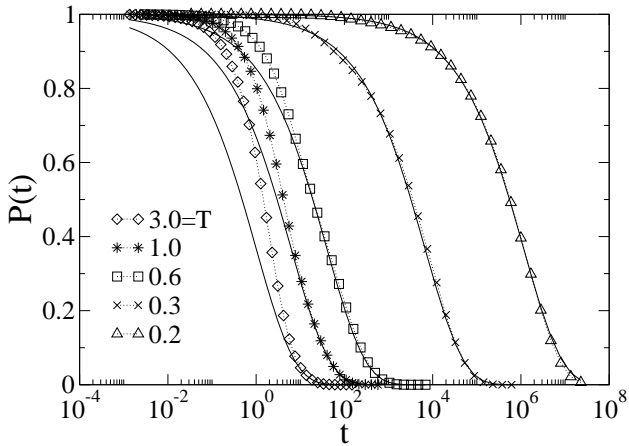,width=8.25cm}
\caption{\label{persistence} Persistence functions in the 1D FA model
for the same set of temperatures as in Fig.\ \ref{distrib}.  Symbols
are numerical data and full lines are fits to the stretched
exponential form expected theoretically for low temperatures, $P(t) =
\exp \left[ -(t/\tau(T))^\beta \right]$, with $\beta=1/2$.}
\end{figure}

From the distributions of times, Fig.\ \ref{distrib}, it is easy to
derive dynamical correlators for the 1D FA model considered here.  The
spatially averaged persistence function reads,
\begin{equation}
P(t) = \int_t^\infty \upd \tau' \pi(\tau').
\end{equation} 
The behaviour of $P(t)$ as a function of time for various temperatures
is shown in Fig.\ \ref{persistence}.  At very low temperatures, the
persistence function is known exactly due to the diffusion properties
of isolated defects, and one gets
\begin{equation}
P(t) = \exp \left[ - \left( \frac{t}{\tau(T)} \right)^{\beta} \right],
\label{stretched}
\end{equation}
where $\tau(T)$ is the relaxation time discussed in the next section.
For the 1D FA model, $\beta=1/2$, 
but the stretching exponent 
might be temperature dependent in more elaborated
(fragile) models~\cite{Berthier-Garrahan,Ritort-Sollich}, as is indeed
observed in experiments~\cite{Ediger-et-al,Angell}.

We see from Fig.\ \ref{persistence} that for $T=0.2$ and $0.3$
$P(t)$ is well approximated by Eq.\ (\ref{stretched})
on the whole time window.
For higher temperatures, $T > T_c = 0.3$, the mixed character of the
correlators is evident from the fact that Eq.~(\ref{stretched}) only
describes the long time behaviour of the correlator, as expected.  In
this temperature regime, short times are best described by a simple
exponential.  In the high temperature regime, $T > T_o =1.0$,
relaxation is just exponential for all times.  We conclude that the
appearance of isolated defects at $T_o$ is reflected in the long time
behaviour of dynamical correlators.  In the crossover region, $T_c < T
< T_o$, more and more of the decorrelation is due to isolated defects
when $T$ decreases. Below $T_c$ the entire decorrelation 
is due to these slow processes.

A confirmation of the progressive domination of slow over fast
processes described above can be found in the numerical results of
Ref.\ \cite{schroder}.  The similarity of Fig.\ 4 of Ref.\
\cite{schroder} and our Fig.\ \ref{persistence} is 
in fact quite striking.  In
particular, Ref.\ \cite{schroder} calculated density-density
correlations from both real configurations and their corresponding IS
in a binary Lennard-Jones mixture.  The latter, where thermal energies
were removed in the quenching procedure, are the ones which have to be
compared with Fig.\ \ref{persistence}, since 
fast vibrations are also removed in our coarse-grained approach.

An important conclusion is that the long-time decay of dynamical
correlators, sometimes referred to as the alpha-relaxation, is due to
the presence of isolated defects and therefore of heterogeneous
dynamics, even in the $T > T_c$ regime.  
This means that activated dynamics, in the language of MCT, 
or hopping events in topographic terms, are responsible for
the alpha-relaxation, even at temperatures well above $T_c$.
This conclusion is unavoidable in view of the numerical data of 
Refs.~\cite{schroder,Doliwa-Heuer,Denny-et-al}.  

\subsection{Relaxation time}

\begin{figure}
\epsfig{file=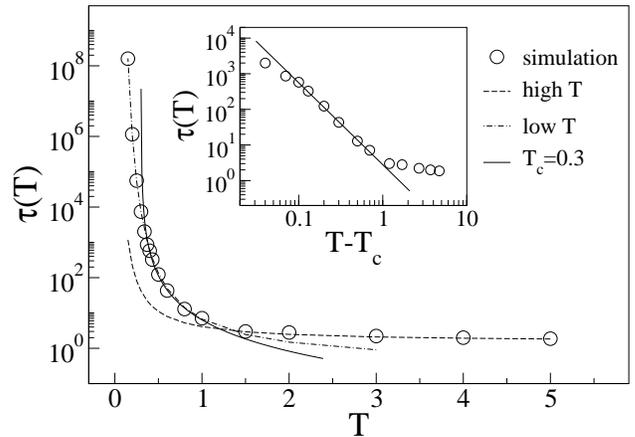,width=8.25cm}
\caption{\label{talpha} Temperature dependence of the relaxation time
in the 1D FA model, $\tau(T)$. Open circles correspond to numerical
data.  Three fits are presented.  The dashed line is the simple
mean-field, Hartree like approximation, $\tau_{\rm MF} \sim \exp(1/T)$.
The dotted-dashed line is the low-$T$ exact behaviour, $\tau_{\rm ex} \sim
\exp(3/T)$.  The full line is a power law MCT-like fit, $\tau_{\rm MCT}
\sim (T-T_c)^{-\gamma}$ with $\gamma =2.3$ and $T_c =0.3$.  The inset
shows that the apparent power law behaviour is acceptable in a range
of three decades in times.  The main figure shows that low-$T$ and
high-$T$ fits account for the whole temperature range.}
\end{figure}

The next natural quantity to consider, the relaxation time $\tau(T)$,
is readily obtained from the dynamical correlators discussed in the
previous subsection.  From the discussion of section~\ref{physics}, we
expect a crossover from high-$T$ to low-$T$ at the onset temperature
$T_o$, see Eq.\ (\ref{To}).  Our results for the 1D FA model are
presented in Fig.\ \ref{talpha}, where $\tau(T)$ is defined as the
time where the persistence function has decayed to the value $1/e$.

The simplest mean-field approximation to the dynamics of the FA model
consists in a Hartree like decoupling of spatial correlations,
$\langle n_i n_j \rangle \to n_i \langle n_j \rangle$, in the
dynamical equation for $n_i$.  This amounts to replacing the actual
neighbourhood of site $i$ by an average neighbourhood, and spins are
always facilitated with an average rate equal to $c$.  This gives a
mean-field estimate of the relaxation time \cite{schulz},
\begin{equation}
\tau_{\rm MF}(T) \approx c^{-1} \sim \exp \left( \frac{1}{T} \right) .
\end{equation}
Fig.\ \ref{talpha} shows that this simple approximation accounts for
the dependence of the relaxation time at high temperatures,
$T \ge T_o$.

The exact result for the relaxation time of the 1D FA model is obtained by
realizing that isolated 
defects undergo diffusion with a temperature dependent
diffusion constant, $D(T) \approx c \sim \exp(-1/T)$.  The system
relaxes when defects have diffused over a distance given by the mean
separation between defects, $c^{-1}$, so that
\begin{equation}
\tau_{\rm ex}(T) \approx D^{-1} c^{-2} \sim \exp \left(
\frac{3}{T} \right).
\label{tex}
\end{equation}
This mechanism relies on the notion of dynamic
facilitation which implies that local fluctuations of the mobility
determine the dynamics, and is essentially beyond
reach of any mean-field type of approximation \cite{Harrowell}.  We
see from Fig.\ \ref{talpha} that Eq.\ (\ref{tex}) accounts for the
behaviour at low temperatures, $T \le T_o$.
Figure \ref{talpha} also presents a
fit to the data with an MCT power law form for the relaxation time
\cite{MCT},
\begin{equation}
\tau_{\rm MCT}(T) \approx (T-T_c)^{-\gamma} ,
\label{tmct}
\end{equation}
similar to the one obtained in Ref.\ \cite{Fredrickson-Andersen} for
the 2-spin facilitated, two dimensional version of the FA model.

From Fig.\ \ref{talpha}, we draw the following conclusions.  The
behaviour of the relaxation time changes from the high-$T$ to low-$T$
behaviour close to the onset temperature $T_o$.  The combination of
simple mean-field at high-$T$ with the exact form at low-$T$ allows to
describe the temperature dependence of the relaxation time over the
whole temperature range.  However, given that $\tau(T)$ smoothly
interpolates between these two different functional forms, the MCT power
law form, Eq.\ (\ref{tmct}), appears to works
reasonably well in a time window of about three decades (see inset in
Fig.\ \ref{talpha}).  This range of apparent power-law behaviour is in
fact larger than the corresponding one in the canonical binary
Lennard-Jones mixture of Ref.~\cite{Kob-Andersen}, where extensive
tests of MCT have been performed.  Remarkably, we also find that the
$T_c$ extracted from the power law fit to the relaxation time
coincides well with the temperature where fast processes
cease to contribute in a significant manner to the distribution of
relaxation times, Fig.\ \ref{distrib}.  This explains our choice of
notation for the lower crossover temperature, $T_c$.

Following the standard MCT reading of the 
data~\cite{MCT-simulations,Kob-Andersen,MCT-experiments}, we would
erroneously conclude that activated processes only appear close to $T_c
= 0.3$, since these processes are often tautologically defined by the
breakdown of the power law behaviour of the relaxation time
\cite{MCT-simulations}.  Figs.\ \ref{distrib}, \ref{persistence}, and
\ref{talpha}, prove instead that activated dynamics starts to be
relevant at $T_o$, much above $T_c$, dominating the alpha-relaxation
of the correlators, and hence the relaxation time of the system.
The results of this section considerably
weaken the possibility of the existence of a temperature regime in
supercooled liquids where the relaxation time is correctly described by a
power law behaviour.  
It follows that the standard determination of the location
of the MCT ``singularity'' $T_c$ in 
experiments and simulations is physically unjustified~\cite{tarjus,tarjus2}.
In fact, the issue of the location of $T_c$ has been recently addressed
in Ref.~\cite{reichman}, where it was found that for a variety of systems,
the temperature $T_c$ obtained from the actual MCT equations 
systematically coincides
with the onset temperature $T_o$ discussed above.

\subsection{Crossover temperatures}

\begin{figure}
\epsfig{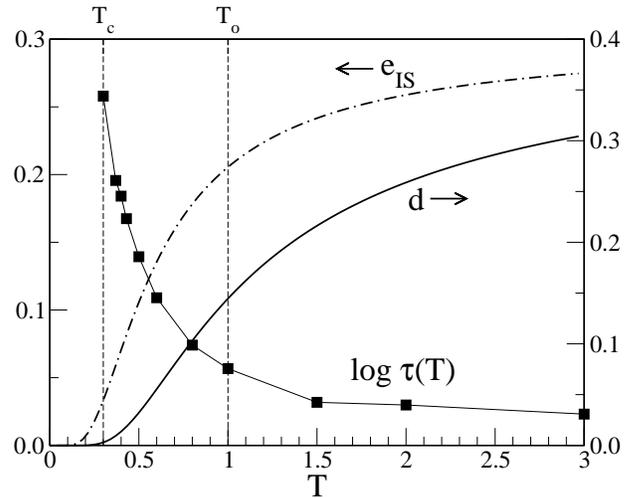}
\caption{\label{crossoverFA} Onset of the slowing down in the 1D FA
model.  We show three quantities as a function of temperature $T$.
(i) Logarithm of the relaxation time, $\log \tau$, see Fig.\
\ref{talpha}.  (ii) energy of IS, $e_{IS}$, which displays a
qualitative change around $T_o = 1.0$; (iii) concentration $d$ of
cells moved in the descent from an equilibrium configuration to its
IS, which displays a qualitative change around $T_c=0.3$.  This figure
should be compared with Fig.\ \ref{crossoverLJ}.}
\end{figure}

Let us now consider the quantities shown in Fig.\ \ref{crossoverLJ} as
evidence in favour of the MCT/landscape picture in Lennard-Jones
mixtures, from the perspective of dynamically facilitated models.  In
Fig.\ \ref{crossoverFA} we show for the 1D FA model a plot analogous
to Fig.\ \ref{crossoverLJ}.   

The first quantity presented in Fig.\ \ref{crossoverFA} is the
logarithm of the relaxation time, $\log \tau(T)$, as a function of
temperature, which we discussed in detail in the previous subsection.

The second quantity shown in Fig.\ \ref{crossoverFA} is the average
energy of inherent structures, $e_{IS}$.  For the 1D FA model it can
be computed analytically by solving the zero-temperature dynamics of
the model \cite{ritort},
\begin{equation}
e_{IS}(T) = c ~ e^{-c},
\label{eis}
\end{equation}
where the concentration of defects, $c$, is defined in 
Eq.~(\ref{ct}).  At high temperature $e_{IS}$ changes very slowly.  When
$T$ is reduced below $T_o$ the concentration of defects starts to
decrease markedly, and $e_{IS}$ follows the same trend, as can be seen
in Fig.\ \ref{crossoverFA}.  This change in behaviour at the onset
temperature $T_o$ is due to the appearance of isolated
defects, and therefore of a heterogeneous dynamics, and not to any
special change of the potential energy surface.  This is a very
different interpretation of the physics from that of Ref.\
\cite{Sastry-et-al}.

The third quantity shown in Fig.\ \ref{crossoverFA} is analogous to
the distance between a configuration and its nearest IS (see Fig.\
\ref{crossoverLJ}).  In the lattice models we are considering the
natural quantity to compute is the concentration of sites which change
during the descent towards the inherent structure, $d(T)$.  Since only
excited sites can change during this procedure, we get,
\begin{equation}
d(T) = c - e_{IS} = c \left( 1- e^{-c} \right).
\end{equation}
Clearly, at low temperature $d \approx c^2 \sim \exp(-2/T)$. This
behaviour is physically natural.  Contributions to $d(T)$ come from
clusters of defects, which are the only objects that can relax during
the descent to an IS.  Since the probability for a cluster of $k$
defects, $p(k)$, goes as $p(k) \approx c^k$, the main non trivial
contribution to $d(T)$ at low $T$ comes from the smallest clusters,
$k=2$.  These relax only one defect in the descent, so that $d \approx
p(2) \approx c^2$.  Moreover, our defect interpretation is consistent
with real space observations in simulations of silica \cite{jund},
where it was found that during the descent to the IS the major
contributions to the distance comes from annihilation of 
localized topological
defects of the amorphous structure.

The similarity of Fig.\ \ref{crossoverFA} to Fig.\ \ref{crossoverLJ}
is striking.  The emerging physical picture is however completely
different from the one of the MCT/landscape scenario.  For example,
while it may appear from the behaviour of $d(t)$ above $T_c$ that this
quantity extrapolates to zero when $T \to T_c$ (see Fig.\
\ref{crossoverFA}), the exact temperature dependence of $d(T)$ is
purely Arrhenius.  This means that $T_c$ has no particular importance
[$T_c$ would not look special in a plot of $d(T)$ versus $1/T$].  Even
if one accepts $T_c$ as delimiting two regimes with high and low
concentrations of clusters of excitations, this apparent crossover is
completely irrelevant as far as the long-time dynamics is 
concerned. These observations suggest 
that the crossover at $T_c$ from ``landscape-influenced'' to
``landscape-dominated'' of Ref.~\cite{Sastry-et-al} is not physically
significant for the alpha-relaxation.

\section{Interpretation of ``landscape'' properties}
\label{landscape}

In recent years, the potential energy landscape of
supercooled liquids has become an object of study 
{\it per se}~\cite{Debenedetti-Stillinger}.
In Ref.~\cite{Berthier-Garrahan}, 
we have developed the idea that the main motivation
behind these works was the observation
of the separation between fast vibrations and slow hopping processes
if sufficiently small systems are considered.
This apparently harmless statement on the system size, we argued in 
Ref.~\cite{Berthier-Garrahan}, results in fact from the central feature
of the dynamics of supercooled liquids:
``sufficiently small'' really means ``if the system size 
is of the order of the dynamical correlation 
length $\ell(T)$''~\cite{Buchner-Heuer2,Berthier-Garrahan}.
However, in a purely topographic description of the physics based
on the statistical properties of minima, the relevance 
of the dynamical correlation length is not obvious~\cite{ribbon}.
In that sense, a topographic 
description of the glass transition 
misses a central aspect of the physics. 

We shall show below that using the very simple spatial approach
described in previous sections,
we can trivially derive
the statistical properties of the landscape reported in recent years.
This successful confrontation to such an amount of
apparently non-trivial and detailed
numerical results is again
a strong indication of the validity of our approach.

\subsection{Real space description of ``minima'' and ``saddles''} 

\begin{figure}
\epsfig{file=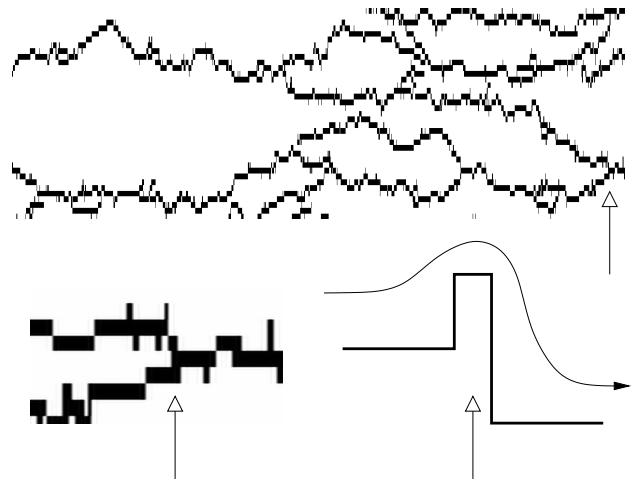,width=8.25cm}
\caption{\label{saddle} Top: zoom on the low-$T$ trajectory of Fig.\
\ref{traj}.  The vertical arrow indicates the closing of a bubble.
Bottom left: expanded view of this event, showing two excitation lines
meeting and coalescing.  A cluster of three spins is needed for this
process to occur.  Bottom right: corresponding ``reaction path''.
Before the event there are two isolated defects (energy $=2$), a
cluster of three defects (energy $=3$) during the event, and one
isolated defect after (energy $=1$).}
\end{figure}

Reference~\cite{Berthier-Garrahan} described in detail the connection
between the nanoscopic ordering in the trajectories of dynamically
facilitated models and the dynamics between IS, ``metabasins'',
or ``traps'' observed in numerical simulations or experiments of
supercooled liquids.  This same approach can be extended to account
for the properties of ``saddles'', i.e., configurations related to
transitions between IS.

Figure~\ref{saddle} zooms on the lowest temperature trajectory of
Fig.~\ref{traj}.  The top panel of Fig.\ \ref{saddle} shows diffusion
of isolated defects.  It also shows coalescence and branching events,
i.e., closing and opening of bubbles.  These two kinds of processes
correspond to ``hopping'' events between dynamical ``traps''
\cite{Berthier-Garrahan}.  Lets consider in detail one of these
events, for example the coalescence process enlarged in the bottom
left panel of Fig.\ \ref{saddle}.  Isolated defects diffuse by first
facilitating one of their neighbours, for instance,
\begin{equation}
10 \to 11 \to 01 .
\label{diff}
\end{equation}
For two defects to coalesce the minimum number of excitations that
have to be present when they merge is three:
\begin{equation}
101 \to 111 \to 011 \to 010 .
\label{trans}
\end{equation}
In this sequence, the total number of defects is 2 at the beginning, 3
at the transition, and 1 at the end.  This process is schematically
described in the bottom right panel of Fig.\ \ref{saddle}.

In topographic terms, an isolated defect corresponds to a local
minimum of the energy, since such a configuration can only evolve
by an energy increase, as in (\ref{diff}).  On the other hand, a
cluster of three excitations corresponds locally to a saddle
point, since it is the transition configuration between two
minima, as in (\ref{trans}) and Fig.~\ref{saddle}.  Larger
clusters thus correspond to higher order saddles, since
the larger the cluster the larger the number of possible moves
into minima. The case $k=2$ is
particular: it is not a minimum since it can relax one defect to
decrease its energy, but it is not a saddle either since it does
not correspond to a hopping event like that of Fig.\ \ref{saddle}.
Clusters with $k=2$ are just ordinary points (i.e., not stationary
points) of the landscape.  The previous discussion generalizes in a
natural way to the whole class of dynamic facilitated systems.

\subsection{Absence of ``geometric transition''}
\label{geometric}

The above identification between the relevant dynamical objects,
isolated defects and clusters of defects, and ``landscape properties''
allows one to compute quantities such as the mean saddle index,
$n_s(T)$, and the corresponding mean energy of stationary points,
$e_s(T)$.

The quantities $n_s$ and $e_s$ were estimated numerically in
simulations of supercooled liquids
\cite{Angelani-et-al,Broderix-et-al}.  It was found that both
functions decrease when $T$ decreases, and extrapolations were
performed that indicated $n_s(T_c)=0$.  Also, plotting the dependence
of $n_s$ on $e_s - e_{IS}$ parameterized by the temperature, a
simple linear relation was obtained, 
$n_s \propto (e_s - e_{IS})$~\cite{Angelani-et-al,Broderix-et-al,saddle2}.

\begin{figure}
\epsfig{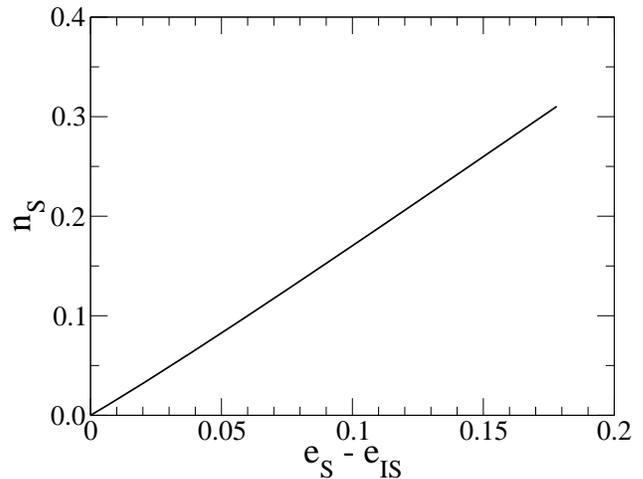}
\caption{\label{geometric2} Saddle index versus energy difference
between saddle and minima computed analytically for 
the 1D FA model for $T\in [0,\infty)$ and $p_s=1/2$.
The obtained linear behaviour is a natural
consequence of (i) dynamic facilitation, (ii) localized defects
and (iii) dynamic heterogeneity.}
\end{figure}

In the case of the 1D FA model, it is very simple to devise a
procedure to go from an equilibrium configuration to the ``nearest''
stationary point.  Isolated defects and clusters of defects with $k
\ge 3$ are locally such points, so we only have to deal with $k=2$
clusters.  From these we can either reach a ``minimum'' $k=1$ or a
saddle with $k=3$.  We 
respectively assign the probabilities $p_s$ and
$(1-p_s)$ to these two possibilities.  We then have,
\begin{equation}
n_s (T) = \sum_{k=1}^\infty p(k) n_s(k),
\end{equation}
and 
\begin{equation}
e_s (T) = \sum_{k=1}^\infty p(k) e_s(k),
\end{equation}
where $p(k)=(1-c)^2 c^k$ is the probability to have a cluster of size
$k$.  From the discussion above we know that $n_s(1)=0$, $e_s(1) = 1$,
$n_s(2) = (1-p_s)$, $e_s(2) = p_s + 3(1-p_s)$, 
$n_s(k \ge 3) = e_s(k \ge 3) = k$.  
Putting all together we obtain,
\begin{equation}
n_s(T) = 3 c^2 \left[ (1-p_s) (1-c)^2 + c \left( 1- \frac{2}{3} c
\right) \right],
\label{index}
\end{equation}
and
\begin{equation}
e_s(T)  = c + (1-2 p_s) c^2 (1-c)^2 .
\label{ens}
\end{equation}
At low $T$ both quantities scale as $n_s \sim c^2$ and $e_s \sim c$.
Three important conclusions can be drawn: 

(i) It is obvious from Eq.\ (\ref{index}) that $n_s(T) > 0$ for
$T>0$. This means that there is no ``geometric transition'' to a
regime with vanishing saddle index.

(ii) The saddle index has a temperature dependence which follows
closely that of the distance $d(T)$ discussed in the previous
subsection.  This is expected because they both receive their
principal contribution, at low $T$, from clusters with $k=2$.  In
other words, the main objects for the low temperature dynamics, the
isolated defects, do not contribute to these quantities.  Therefore,
as for the distance $d(T)$ in Fig.\ \ref{crossoverFA}, the rapid
decrease of $n_s(T)$ when the temperature decreases can easily be
confused with a vanishing of the saddle index close to $T_c$.

(iii) The linear relation between $n_s$ and $e_s - e_{IS}$ becomes
exact at low temperatures, see Eqs.\
(\ref{eis},\ref{index},\ref{ens}).  Again, the difference between
$e_s$ and $e_{IS}$ comes from the clusters of defects which are
relaxed during the descent to the inherent structure.  In Fig.\
\ref{geometric2}, we show the behaviour of $n_s$ versus $e_s - e_{IS}$
for the entire range $T \in [0,\infty)$.  Note that in Ref.\
\cite{Broderix-et-al} a linear behaviour between $n_s$
and $e_s$ was also reported.  This is true at relatively high
temperature, given that above $T_o$ the energy of inherent structure
is almost constant while $n_s$ and $e_s$ change with temperature in the
same way.

These results are valid beyond the FA model which we have used to
illustrate them.  The inexistence of the geometric transition where
the saddle index vanishes follows from the observation that the low
temperature behaviour of $n_s$ is given by the smallest cluster of
defects necessary to make a transition, in the sense described in the
previous section.  Since these objects are spatially localized, their
energy cost is $O(1)$, and they exist with non zero probability at
finite temperature, $T>0$.  This argument is close in spirit to
Stillinger's argument for the inexistence of an entropy crisis at the
Kauzmann temperature involving point defects \cite{still}.  Moreover,
as discussed in the introduction, careful numerical simulations both
confirm that $n_s(T<T_c) >0$ \cite{doye,stariolo}, 
and report an Arrhenius
behaviour $n_s(T)$ \cite{dh2}, in agreement with our results.

The 
relation between saddle index and energy,
\begin{equation}
n_s \propto (e_s - e_{IS}),
\label{lin}
\end{equation}
which was first observed numerically
\cite{Angelani-et-al,Broderix-et-al,saddle2}, is also a general result for
dynamically facilitated systems.  This relation contains two different
pieces of information.  First, it shows that the intensive saddle
index is a number of $O(1)$.  This is again a trivial consequence of
the existence of the dynamical correlation length $\ell(T)$, so that a
large sample can in fact be thought of an assembly of independent
subsystems of linear size $\ell(T)$ \cite{Berthier-Garrahan}.  Second,
and more interesting, is a connection between energy and saddle index,
presented as a ``general feature of the
potential energy landscape of supercooled liquids'' \cite{saddle2},
for which no theoretical explanation was 
however available.  This feature is
in fact almost a tautology in the context of facilitated models: the more
defects are present, the more available directions to move, the
higher the energy above that of the IS. The fact that relation
(\ref{lin}) holds in different model liquids is another confirmation
that dynamical facilitation is a key generic feature of the dynamics
of supercooled liquids.

\subsection{Thermodynamics and ``anharmonicities''}
\label{SW}

Another common procedure of the landscape approach is to decompose
configurations into vibrational and configurational components.
Stillinger and Weber~\cite{Stillinger} suggested to perform this
decomposition at the level of the partition function,
\begin{equation}
Z(T) \approx \sum_{E_{IS}} \Omega(E_{IS}) \exp \left( - \frac{E_{IS} +
F(T;E_{IS})}{T} \right),
\end{equation}
where the sum is over energies of IS, $E_{IS}$, their number is
indicated by $\Omega(E_{IS})$, and $F(T,E_{IS})$ is the ``basin free
energy'' which takes into account fluctuations within an IS due to
vibrations and possible ``anharmonicities'' (i.e., all the rest).

It is instructive to consider the calculation of the partition
function in the case of the 1D FA model using the Stillinger
and Weber decomposition. The
thermodynamics of the FA model is that of a non-interacting gas of
binary excitations.  This simple thermodynamics, however, can be
obtained with any dynamics obeying detailed balance with
respect to the Hamiltonian~(\ref{ham}), the actual FA
dynamics defined by (\ref{rates}) being just one possibility.
In this sense, the Stillinger and Weber 
prescription for thermodynamics is an approximation for the way a
thermodynamic quantity would be calculated using a particular choice
of dynamics to sample the configuration space. Inherent
structures, basin free energies, etc., have no thermodynamic meaning,
they only have a dynamical meaning associated with a particular
choice of dynamics.

For the 1D FA model we can evaluate the Stillinger
and Weber partition function exactly.
We have:
\begin{eqnarray}
Z_N(T) &=& \sum_{E_{IS}=0}^{N/2} \frac{\left(N-E_{IS}\right)!}{E_{IS}!
~ \left(N-2E_{IS}\right)!} \nonumber \\ && \times
\exp{\left(-\frac{1}{T} \left[ E_{IS} + F_{\rm anh}(T;E_{IS}) \right]
\right)} .
\label{ZN}
\end{eqnarray}
The first factor counts the number of configurations of energy
$E_{IS}$ with only isolated defects in a system of $N$ sites.  Due to the
coarse-grained nature of facilitated models the only contribution
to the basin free energy comes from anharmonicities.

Performing the sum in Eq.\ (\ref{ZN}) without the anharmonic
contribution gives,
\begin{equation}
Z_N^{\text{harm}}(T) = \exp \left(- \frac{N}{2T} \right) U_N\left[
-\frac{i}{2} \exp{\left(\frac{1}{2T} \right)} \right] ,
\end{equation}
where $U_N(x)$ is the $N$-th Chebyshev polynomial of second kind.  In
the thermodynamic limit the above expression simplifies to give the
free energy in the ``harmonic'' approximation,
\begin{eqnarray}
f_{\text{harm}}(T) &=& \lim_{N \to \infty} -\frac{T}{N} \ln Z_N(T)
\nonumber \\ &=& T ~ \ln{2} - T ~ \ln{\left( 1 + \sqrt{1+4~e^{-1/T}}
\right)}.
\label{fharm}
\end{eqnarray}

\begin{figure}
\epsfig{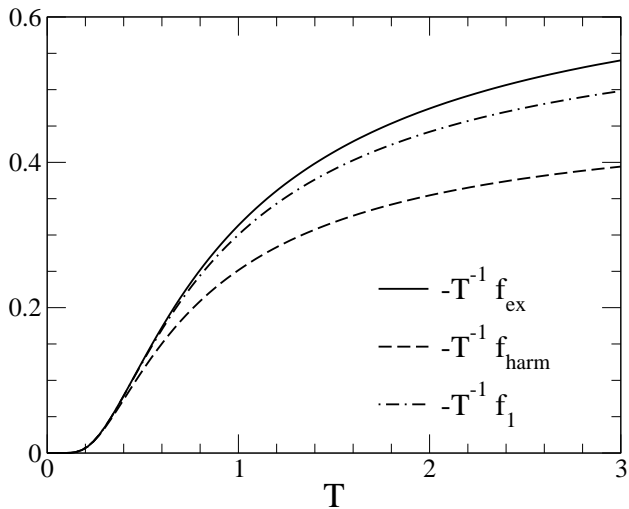}
\caption{\label{figSW} Comparison of the various expressions for the
free energy (divided for convenience by $-T$) in the 1D FA
model: $f_{\rm ex}$ is the exact free energy (\ref{fex}); $f_{\rm
harm}$ is the purely harmonic evaluation (\ref{fharm}) which is a good
approximation below $T_c=0.3$, and $f_1$ is the result obtained with
the expression (\ref{approx1}) for anharmonicities which is good up to
$T_o=1.0$.}
\end{figure}

In Fig.\ \ref{figSW}, we compare the approximation (\ref{fharm}) to
the exact expression for the free energy,
\begin{equation}
f_{\rm ex}(T) = - T \ln{\left( 1 + e^{-1/T} \right)} .
\label{fex}
\end{equation}
As observed numerically in supercooled liquids
\cite{Debenedetti-Stillinger}, both thermodynamic evaluations,
Eqs. (\ref{fharm}) and (\ref{fex}), apparently coincide below $T_c$
when anharmonicities become negligible. From the previous sections
we know that anharmonicities are just a consequence of the existence
of clusters of defects.  At low temperature the difference between
(\ref{fharm}) and (\ref{fex}) is therefore 
proportional to $c^2$,
reflecting the
fact that: (i) anharmonicities do not disappear below $T_c$, which
again is no particular temperature in this context; (ii)
anharmonicities are due to clusters of defects, $k=2$ being the
leading term at low temperatures.

In the particular case of the FA model we can formulate an exact
expression for the anharmonic free energy $F_{\rm anh}$.  It is easy to
check that the choice
\begin{equation}
F_{\rm anh}(T;E_{IS}) = E_{IS}(T) ~ f_{\rm ex}(T)
\label{fanh}
\end{equation}
in Eq.\ (\ref{ZN}) yields the exact expression for the free energy in
the thermodynamic limit.  This exact expression for the
anharmonic part of the free energy is simple to understand.  In an IS
all defects are isolated.  The contribution of clusters of defects is
obtained from the probability that an isolated defect in the IS was a
cluster in the original configuration, which is given by Eq.\
(\ref{fanh}).

Interestingly, a numerical procedure to evaluate the anharmonic
contributions can be proposed.  First, the harmonic expression
(\ref{fharm}) is evaluated.  Then, the approximate expression
\begin{equation}
F_{\rm anh}(T;E_{IS}) \sim E_{IS}(T) ~ f_{\rm harm}(T)
\label{approx1}
\end{equation}
can be used as an educated guess for the anharmonic contributions.  This
gives in turn 
a first order free energy, $f_1(T)$.  The improvement on the
harmonic evaluation can be judged in Fig.\ \ref{figSW}, where we see
that $f_1(T)$ coincides with the exact free energy up to $T \sim T_o$.
This evaluation can then be improved iteratively using $F_{\rm
anh}(T;E_{IS}) \sim E_{IS}(T) ~ f_1(T)$ to get $f_2(T)$, and so on.
It is easy to show that, 
in our particular case, $\lim_{n \to \infty} f_n(T) = f_{\rm ex}(T)$.

Although these results could lead to an improvement on present 
evaluations of anharmonic contributions in studies of the thermodynamics
of supercooled liquids, they also show 
that topographic concepts are very far from the
physical objects they pretend to describe.

\subsection{Failure of Adams-Gibbs relation}

We end this section with a remark on the Adam-Gibbs relation, which is
an attempt to connect dynamical properties to thermodynamic ones. The
Adam-Gibbs formula relates the relaxation time of a glass former,
$\tau_\alpha$, to the configurational entropy, $S_c$ [which would
correspond to $S_c = \langle \ln{\Omega(E_{IS})} \rangle$ in the IS
formalism]: $\tau_\alpha \propto \exp \left[ 1/(T S_c) \right]$.
Apparently, this relation has been seen to hold both in numerical
simulations and in experiments of various systems
\cite{Debenedetti-Stillinger}.  
A careful look at the published data reveals however
that the correlation between relaxation time and entropy 
does not quantitatively satisfy the 
Adams-Gibbs relation. 
This important observation is often not clearly 
stated~\cite{Debenedetti-Stillinger}.

Our analysis shows indeed that
thermodynamic properties do not fully determine dynamical behaviours.
Clearly, almost by definition, $\tau_\alpha$ increases and $S_c$
decreases as temperature is lowered, but that is where the connection
ends. It is easy to check that the Adam-Gibbs formula fails
completely when applied to dynamic facilitated systems. 
We find instead that time scales are broadly distributed, 
the distribution
of times being the result of an integral over a
distribution of length scales scale, $\rho(\ell)$, 
imposed by thermodynamic equilibrium, 
Eq.~(\ref{ham}). Crucially however, dynamics 
also enters the integral in the form
of the conditional probability of time and length, $\rho(t|\ell)$, 
which can be described, in a topographic language as 
containing informations on the relevant ``barriers'', 
which have a priori no obvious link with the statistics of minima.
As a consequence, thermodynamics alone cannot be used to predict
the dynamical behaviour. 
Again, we find in the literature an excellent
numerical confirmation of this statement. 
In Refs~\cite{dh2,heuer}, using a purely topographic
description of a supercooled liquid, it was
shown that the diffusion constant
could be computed by a combination of thermodynamic {\it and} dynamical
quantities, well in line with the above discussion. 

\section{Conclusions} 
\label{conclusion}

In this paper, we have developed a spatial description of the physics
of the progressive slowing down of supercooled liquids.  The only
ingredients in our method have been the notions of localized mobility
excitations and facilitated dynamics
\cite{Garrahan-Chandler,Berthier-Garrahan,Garrahan-Chandler2,Berthier}.  
Our results were illustrated explicitly for the simplest case of the 1D FA
model, but are generic for this
theoretical approach, and are in very good agreement with experimental and
numerical observations in supercooled liquids.

The physical picture which emerges from our work is 
however markedly different
from that of the MCT/landscape scenario discussed in the
Introduction. 

At high temperatures, $T > T_o$, the dynamics is fast and
liquid-like, corresponding to the relaxation of large clusters of
defects.  Dynamic facilitation plays no major role, and a simple
mean-field Hartree-like decoupling of the equations of motion yields
predictions in good agreement with numerical results.

When $T < T_o$, the dynamics becomes heterogeneous, in the
sense that local relaxation times are spatially correlated in a non
trivial way.  This can be seen in the trajectories of Fig.\ \ref{traj}
as the appearance of slow 
bubbles~\cite{Garrahan-Chandler,Berthier-Garrahan}.  The
long-time dynamics of the system results from the wide joint
distribution of length scales and time scales, and the relaxation
becomes stretched.  This dynamic heterogeneity, which can be thought
of as the activated dynamics invoked, but never described, by MCT,
determines the alpha-relaxation and its temperature
dependence for $T<T_o$.  Also, dynamic heterogeneity implies that
decoupling of transport coefficients actually starts at $T_o$, as
confirmed by the simulations.  From a theoretical point of view, local
fluctuations of mobility crucially influence the dynamical behaviour.
Any mean-field like approach, no matter how involved,
is most probably doomed to fail.

At $T_o$ not all trace of high-$T$ physics (clusters of
defects) disappears.  The dynamics has a mixed character in the range
$T_c < T < T_o$, as seen for example in the behaviour of dynamical
correlators like in Fig.\ \ref{persistence}.  The temperature $T_c$
is just a crossover.  It is the temperature below which isolated
defects not only dominate the long time dynamics (as for $T_c < T <
T_o$) but are also the most numerous dynamical objects, see 
Eq.~(\ref{Tc}).  Clusters of defects, whose dynamics is homogeneous and
non-activated, are responsible for the temperature dependence of
several quantities, such as distance between configurations and IS,
saddle index $n_s(T)$, and anharmonic contributions to the free
energy.  In numerical simulations, $T_c$ has been interpreted as a key
temperature, in accordance with MCT for which it represents a
dynamical singularity.  We have shown, however, that all of these
quantities have a smooth temperature dependence, 
as has been recently
observed numerically \cite{doye,stariolo,dh2}. This means that $T_c$ does not
correspond to a transition or singular point, 
but is at most a crossover.
Crucially, the objects which display a crossover close to $T_c$ 
are also irrelevant for the
long-time dynamics, so that the inexistence of the singularity $T_c$
is anyway not a physically important issue for the alpha-relaxation.

Below $T_c$, isolated defects are the only remaining objects
and the dynamics is dominated by the nanoscopic demixing of slow and
fast regions so that trajectories look like a dense
mixture of slow bubbles, which in turn gives a natural theoretical
interpretation of the canonical features of glass transition 
phenomena~\cite{Garrahan-Chandler,Berthier-Garrahan,Garrahan-Chandler2}.

Our results, together with some other recent 
studies~\cite{dh2,Doliwa-Heuer,Denny-et-al,gilles1,Harrowell,heuer,tarjus,tarjus2,reichman,Kivelson-et-al}, 
suggest that several essential features of the dynamics of
supercooled liquids need to be recognized and we now list some of them.

\begin{itemize}

\item
The dynamics is
heterogeneous and activated well above $T_c$.  

\item The dynamical
slowing-down of supercooled liquids is due to the growth, below $T_o$,
of a dynamic correlation length $\ell(T)$, or more precisely, of a
whole distribution of length  and time scales.  

\item
The long-time dynamics, and therefore the
relaxation time $\tau_\alpha$ of the
liquid is dominated by heterogeneous ``activated'' dynamics
below $T_o$. 

\item
The MCT definition of activated processes 
as deviations from the ideal theory is incorrect.
It is unlikely that the power law behaviour predicted by MCT
correctly describes the temperature dependence of $\tau_\alpha$.
The practical definition of the temperature $T_c$ 
cannot be used.

\item No topological change of the potential energy landscape
takes place close to $T_c$.  Quantities such as the saddle index and
anharmonicities do not vanish close to $T_c$
and have a smooth temperature behaviour. At best, they undergo a
crossover from large to small which remains to be quantified.

\item
Even if one accepts $T_c$
as a crossover temperature, as in Eq.\ (\ref{Tc}), quantities related
to this crossover are unimportant for the long-time dynamics. 

\item
Knowledge of thermodynamic properties is not
enough to predict dynamical behaviour, which explains the quantitative
failure of relations like the Adams-Gibbs formula.

\end{itemize}

The approach we developed in this paper, which is an extension of
previous efforts
\cite{Garrahan-Chandler,Berthier-Garrahan,Garrahan-Chandler2,Berthier}, is
generic.  It can be applied both to systems like Lennard-Jones liquids
or to hard sphere systems.  It gives a perspective on the physics of
glass formers which is clearly distinct to, and in many respects much
more natural than, that of MCT or topographic approaches.

There are many important and interesting open questions which need to
be addressed from this new perspective.  This include,
among others, understanding properly the origin of mobility
excitations, and
the breakdown of Stokes-Einstein-Debye relations and
associated decouplings between transport coefficients.  

A general
conclusion that can be drawn from this work and our previous ones is
that, in many respects, glass transition phenomena is more standard
than often assumed, in the sense that it is determined by the
interplay between growing dynamic lengthscales and timescales.
This is obviously reminiscent of critical phenomena~\cite{Berthier},
meaning that it should be possible to adapt 
renormalization group techniques 
to study the dynamics of the glass transition.

\begin{acknowledgments}
We are grateful to J.-P. Bouchaud, G. Biroli, A. Buhot, 
A. Heuer, D.R. Reichman, G. Tarjus, and especially
D. Chandler for useful discussions.  We acknowledge financial support
from CNRS (France), EPSRC Grant No.\ GR/R83712/01, EU Marie Curie
Fellowship No.\ HPMF-CT-2002-01927, the Glasstone Fund, and Worcester
College Oxford. Some of the numerical results were obtained on Oswell
at the Oxford Supercomputing Center, Oxford University.  
\end{acknowledgments}


\begin{thebibliography}{99}
\bibitem{Ediger-et-al} M.D. Ediger, C.A. Angell, and S.R. Nagel,
J. Phys. Chem. \textbf{100}, 13200 (1996).

\bibitem{Angell} C.A. Angell, Science \textbf{267}, 1924 (1995).

\bibitem{Debenedetti-Stillinger} P.G. Debenedetti and F.H. Stillinger,
Nature \textbf{410}, 259 (2001).

\bibitem{Debenedetti} P.G. Debenedetti, {\it Metastable liquids}
(Princeton University Press, Princeton, 1996).

\bibitem{Garrahan-Chandler} J.P. Garrahan and D. Chandler, Phys. Rev.
Lett. {\bf 89}, 035704 (2002).

\bibitem{Berthier-Garrahan} L. Berthier and J.P. Garrahan, to appear
in J. Chem. Phys. (2003); cond-mat/0303451.

\bibitem{Garrahan-Chandler2} J.P. Garrahan and D. Chandler,
cond-mat/0301287.

\bibitem{Berthier} L. Berthier, to appear in 
Phys. Rev. Lett. (2003), cond-mat/0303452.

\bibitem{MCT} W. G\"{o}tze and L. Sj\"{o}gren, Rep. Prog. Phys. {\bf
55}, 55 (1992).

\bibitem{MCT-experiments} W. G\"{o}tze, J. Phys. Condens. Matter {\bf
11}, A1 (1999).

\bibitem{MCT-simulations} W. Kob, in {\it Slow relaxations and
nonequilibrium dynamics in condensed matter}, Eds: J.-L. Barrat,
J. Dalibard, M.V. Feigel'man, J. Kurchan (Springer Verlag, Berlin, 2003);
cond-mat/0212344.

\bibitem{DHReviews} For reviews on dynamic heterogeneity, see,
H. Sillescu, J. Non-Cryst. Solids {\bf 243}, 81 (1999); M.D. Ediger,
Annu. Rev. Phys. Chem.  {\bf 51}, 99 (2000).

\bibitem{Kob-Andersen} W. Kob and H.C. Andersen,
Phys. Rev. Lett. {\bf 73}, 1376 (1994);
Phys. Rev. E {\bf 51}, 4626 (1995);
Phys. Rev. E {\bf 52}, 4134 (1995).

\bibitem{thomas} W. G\"otze and T. Voigtmann, Phys. Rev. E {\bf 61}, 
4133 (2000).

\bibitem{kivtar2} G. Tarjus and D. Kivelson, 
in {\it Jamming and Rheology}, Eds: A.J. Liu and S.R. Nagel 
(Taylor and Francis, New York, 2001).

\bibitem{Donth} E.J. Donth, {\it The glass transition} (Springer Verlag,
Berlin, 2001).

\bibitem{Rossler-Sokolov} E. R\"ossler and A.P. Sokolov,
Chem. Geol. {\bf 128}, 143 (1996).

\bibitem{Stickel-et-al} F. Stickel, E.W. Fischer, and
R. Richert, J. Chem. Phys. {\bf 104}, 2043 (1996).

\bibitem{Angell-10Q} 
C.A. Angell, J. Phys. Chem. Sol. {\bf 49}, 863 (1988);
J. Phys. Condens. Matter {\bf 12}, 6463 (2000).

\bibitem{Goldstein} M. Goldstein, J. Chem. Phys. \textbf{51}, 3728 (1969).

\bibitem{Stillinger-Weber} F.H. Stillinger and T.A. Weber,
Phys. Rev. A \textbf{25}, 978 (1982); Phys. Rev. A \textbf{28}, 2408
(1983); Science \textbf{225}, 983 (1984).

\bibitem{Stillinger} F.H. Stillinger, Science \textbf{267}, 1935 (1995).

\bibitem{Sastry-et-al} S. Sastry, P.G. Debenedetti,
and F.H. Stillinger, Nature \textbf{393}, 554 (1998).

\bibitem{Buchner-Heuer1} S. B\"uchner and A. Heuer,
Phys. Rev. E \textbf{60}, 6518 (1999).

\bibitem{Buchner-Heuer2} S. B\"uchner and A. Heuer,
Phys. Rev. Lett. \textbf{84}, 2168 (2000).

\bibitem{schroder} 
T.B. Schroder, S. Sastry, J.C. Dyre, and S.C. Glotzer,
J. Chem. Phys. {\bf 112}, 9834 (2000).

\bibitem{jund} P. Jund and R. Jullien, Phys. Rev. Lett. {\bf 83},
2210 (1999).

\bibitem{keyes1} T. Keyes and J. Chowdhary,
Phys. Rev. E {\bf 65}, 041106 (2002).

\bibitem{boson} For extension of the landscape ideas in the glass
phase see: 
C.A. Angell, Y. Yue, L.M. Wang, J.R.D. Copley, S. Borick, and
S. Mossa, J. Phys. C Condens. Matter {\bf 15}, S1051 (2003);
T. S. Grigera, V. Martin-Mayor, G. Parisi, P. Verrocchio,
Nature {\bf 422}, 289 (2003);
for alternative real space interpretations, see e.g.
E. Duval, L. Saviot, L. David, S. Etienne, J. F. Jal, 
to appear in Europhys. Lett., cond-mat/0306666.

\bibitem{Cavagna} A. Cavagna, 
Europhys. Lett. {\bf 53}, 490 (2001).

\bibitem{Kurchan-Laloux} J. Kurchan and L. Laloux, J. Phys. A {\bf
29}, 1929 (1996).

\bibitem{Cavagna-et-al} A. Cavagna, J.P. Garrahan, and
I. Giardina, Phys. Rev. B {\bf 61}, 3960 (2000).

\bibitem{Broderix-et-al}  K. Broderix, K.K. Bhattacharya, A. Cavagna, 
A. Zippelius, and I. Giardina,
Phys. Rev. Lett. {\bf 85}, 5360 (2000).

\bibitem{Angelani-et-al} L. Angelani, R. Di Leonardo, G. Ruocco, A. Scala, 
and F. Sciortino,
Phys. Rev. Lett. {\bf 85}, 5356 (2000).

\bibitem{saddle2} 
P. Shah and C. Chakravarty, J. Chem. Phys. {\bf 115}, 8784 (2001);
T.S. Grigera, A. Cavagna, I. Giardina, and G. Parisi,
Phys. Rev. Lett. {\bf 88}, 055502 (2002);
L. Angelani, R. Di Leonardo, G. Ruocco, A. Scala, and F. Sciortino,
J. Chem. Phys. {\bf 116}, 10297 (2002);
L. Angelani, G. Ruocco, M. Sampoli, and F. Sciortino,
cond-mat/0305280.

\bibitem{doye} J.P.K. Doye and D.J. Wales, 
J. Chem. Phys. {\bf 116}, 3777 (2002).

\bibitem{stariolo} G. Fabricius and D.A. Stariolo,
Phys. Rev. E {\bf 66}, 031501 (2002).

\bibitem{dh2} B. Doliwa and A. Heuer, Phys. Rev. E {\bf 67}, 031506 (2003).

\bibitem{Cavagna2} A. Cavagna, I. Giardina, and G. Parisi,
J. Phys. A {\bf 34}, 5317 (2001).

\bibitem{laird} S.D. Bembenek and B.B. Laird,
Phys. Rev. Lett. {\bf 74}, 936 (1995);
J. Chem. Phys. {\bf 104}, 5199 (1996).

\bibitem{INM} 
C. Donati, F. Sciortino and P. Tartaglia, Phys. Rev. Lett. {\bf 85},
1464 (2000);
E. La Nave, H.E. Stanley and F. Sciortino, Phys. Rev. Lett. {\bf 88},
035501 (2002).

\bibitem{keyes2} T. Keyes, J. Phys. Chem. {\bf 101}, 2921 (1997).

\bibitem{kobsciortino} 
W. Kob, J.-L. Barrat, F. Sciortino, and P. Tartaglia, 
J. Phys. C {\bf 12}, 6385 (2000).

\bibitem{Doliwa-Heuer} B. Doliwa and A. Heuer, 
Phys. Rev. E \textbf{67}, 030501 (2003).

\bibitem{Denny-et-al} R.A. Denny, D.R. Reichman, and J.-P. Bouchaud,
Phys. Rev. Lett.  \textbf{90}, 025503 (2003).

\bibitem{Glotzer} See for a review, S.C. Glotzer,
J. Non-Cryst. Solids \textbf{274}, 342 (2000).

\bibitem{kivtar3} G. Tarjus and D. Kivelson,
J. Chem. Phys. {\bf 103}, 3071 (1995).

\bibitem{gilles1} P. Viot, G. Tarjus, and D. Kivelson,
J. Chem. Phys. {\bf 112}, 10368 (2000);
D.N. Perera and P. Harrowell, Phys. Rev. E {\bf 54}, 1652 (1996).

\bibitem{Glarum} S.H. Glarum, J. Chem. Phys. {\bf 33}, 639 (1960);
M.C. Phillips, A.J. Barlow, and J. Lamb, Proc. R. Soc. Lond. A {\bf
329}, 193 (1972).

\bibitem{Anderson} P.W. Anderson, in {\it Ill-condensed matter}, 
Eds: R. Balian et al. (North Holland, Amsterdam, 1979).

\bibitem{Palmer-et-al} R.G Palmer, D.L. Stein, E. Abrahams, and
P.W. Anderson, Phys. Rev. Lett. {\bf 53}, 958 (1984).

\bibitem{Fredrickson-Andersen} G.H. Fredrickson and H.C. Andersen,
Phys. Rev. Lett. {\bf 53}, 1244 (1984);
J. Chem. Phys. {\bf 83}, 5822 (1985).

\bibitem{Harrowell} S. Butler and P. Harrowell, J. Chem. Phys. {\bf
95}, 4454 (1991); 
S. Butler and P. Harrowell, J. Chem. Phys. {\bf 95}, 4466 (1991);
P. Harrowell, Phys. Rev. E {\bf 48}, 4359 (1993);
M. Foley and P. Harrowell, J. Chem. Phys. {\bf 98}, 5069 (1993).

\bibitem{Ritort-Sollich} F. Ritort and P. Sollich,
Adv. Phys. {\bf 52}, 219 (2003).

\bibitem{Garrahan-Newman} J.P. Garrahan and M.E.J. Newman,
Phys. Rev. E {\bf 62}, 7670 (2000); J.P. Garrahan,
J. Phys. C Condens. Matter {\bf 14}, 1571 (2002).

\bibitem{silvio} 
S. Franz, C. Donati, G. Parisi, and S.C. Glotzer,
Philos. Mag. B {\bf 79}, 1827 (1999);
S.C. Glotzer, V. Novikov, and T.B. Schroder, J. Chem. Phys. {\bf 112}, 
509 (2000).

\bibitem{doliwa} B. Doliwa and A. Heuer, Phys. Rev. E {\bf 61}, 
6898 (2000).

\bibitem{harro2} 
M.M. Hurley and P. Harrowell, Phys. Rev. E {\bf 52}, 1694 (1995);
D.N. Perera and P. Harrowell,
J. Chem. Phys. {\bf 111}, 5441 (1999).

\bibitem{length} Y. Hiwatari and T. Muranaka, 
J. Non-Cryst. Solids {\bf 235-237}, 19 (1998).

\bibitem{heuer} B. Doliwa and A. Heuer, cond-mat/0306343.

\bibitem{schulz} M. Schulz and S. Trimper,
Phys. Rev. E {\bf 57}, 6398 (1998).

\bibitem{tarjus} X.C. Zeng, D. Kivelson, and G. Tarjus, 
Phys. Rev. E {\bf 50}, 1711 (1994).

\bibitem{tarjus2} X.C. Zeng, D. Kivelson, and G. Tarjus,
Phys. Rev. Lett. {\bf 72}, 1772 (1994).

\bibitem{reichman} Y. Brumer and D.R. Reichman,
cond-mat/0306580.

\bibitem{ritort} A. Crisanti, F. Ritort, A. Rocco, and
M. Sellitto, J. Chem. Phys. {\bf 113}, 10615 (2000).

\bibitem{ribbon} D. Kivelson and G. Tarjus, J. Phys. Chem. B {\bf 105}, 11854
(2001).

\bibitem{still} F.H. Stillinger, J. Chem. Phys. {\bf 88}, 7818 (1988).


\bibitem{Kivelson-et-al} D. Kivelson, S.A. Kivelson, X. Zhao,
Z. Nussinov and G. Tarjus, Physica A \textbf{219}, 27 (1995).


\end{thebibliography}
\end{document}